\documentclass[twocolumn,showpacs]{revtex4}
\usepackage{amsmath}
\usepackage{graphicx}

\newcommand{\eq}{\begin{equation}}
\newcommand{\fine}{\end{equation}}

\begin{document}

\title{Entanglement and Non-locality in a Micro-Macroscopic  system}
\author{Francesco De Martini$^{1,2}$, Fabio Sciarrino $^{3,1}$, and Chiara Vitelli$%
^{1} $}

\begin{abstract}
\end{abstract}

\address{$^{1}$Dipartimento di Fisica dell'Universit\'{a} ''La Sapienza'' and\\
Consorzio Nazionale Interuniversitario per le Scienze Fisiche della Materia,\\
Roma, 00185 Italy\\
$^{2}$ Accademia Nazionale dei Lincei, Roma  \\
$^3$Centro di\ Studi e Ricerche ''Enrico Fermi'', Via Panisperna\\
89/A,Compendio del Viminale, Roma 00184, Italy}

\maketitle

\textbf{In recent years two fundamental aspects of quantum mechanics have
attracted a great deal of interest, namely the investigation on the
irreducible nonlocal properties of Nature implied by quantum entanglement
and the physical realization of the ''Schr\oe dinger Cat''. The last
concept, by applying the nonlocality property to a combination of a
microscopic and of a macroscopic systems, enlightens the concept of the
quantum state, the dynamics of large systems and ventures into the most
intriguing philosophical problem, i.e. the emergence of quantum mechanics in
the real life. Rather surprisingly these two aspects, which appeared in the
same year 1935 by the separated efforts of Albert Einstein, Boris Podolsky
and Nathan Rosen \textrm{(EPR)}\ and of Erwin Schr\oe dinger \cite
{Eins35,Schr35} appear not being generally appreciated for their profound
interconnections\ that establish the basic foundations of modern science.
Likely, this follows from the extreme difficulty of realizing a system which
realizes simultaneously the following three conditions: (a)\ the quantum
superposition of two multiparticle, mutually orthogonal states, call it a
''Macro-system'' (b) the quantum non-separability of this superposition with
a far apart single-particle state, i.e. the ''Micro-system'', (c) the
violation of the Bell's inequality stating that by no means any hidden
variable formulation can simulate the nonlocal correlations affecting the
joint Micro-Macro system \cite{Bell65}.}

\textbf{In the present work these crucial conditions are simultaneously
realized and experimentally tested. A Macro - state consisting of \ }$%
\mathbf{N}\approx 3.5\times 10^{4}$ \textbf{photons in a quantum
superposition and entangled with a far apart single - photon state
is generated. Then, the non-separability of the overall
Micro-Macro system is demonstrated and the corresponding Bell's
inequalities are found to be violated. Precisely, an entangled
photon pair is created by a nonlinear optical process, then one
photon of the pair is injected into an optical parametric
amplifier (OPA)\ operating for any input polarization state, i.e.\
into a phase-covariant cloning machine. Such transformation
establishes a connection between the single photon and the multi
particle fields, as show in Figure 1. We then demonstrate the
non-separability of the bipartite system by adopting a local
filtering technique within a positive operator valued measurement
(POVM) \cite{DeMa05b}. The work shows that the amplification
process applied to a microscopic system is a natural approach to
enlighten the quantum-to-classical transition and to investigate
the persistence of quantum phenomena into the ''classical'' domain
by measurement procedures applied to quantum systems of increasing
size \cite {Zure03}. Furthermore, since the generated Micro-Macro
entangled state is directly accessible at the output of the
apparatus, the\ implementation of significant multi qubit logic
gates for quantum information technology can be  achieved by this
method.\ At last, we demonstrate how our scheme may be up-graded
to an entangled Macro - Macro quantum system, by then establishing
a peculiar nonlocal correlation process between two space-like
separated macroscopic quantum superpositions.}\newline

In recent years quantum entanglement has been demonstrated within a two
photon system \cite{Kwia95}, within a single trapped-ion one-photon system
\cite{Blin04,Volz06}, within a single photon and atomic ensemble \cite
{Mats05,deRi06}, within atomic ensembles \cite
{Juls01,Chou05,Mats06,Lan07,Moeh07} and superconducting qubits \cite{Berk03}%
. In addition, the observation of quantum features has been extended to
''cluster'' entangled states involving four \cite{Pan01}, five \cite{Zhao05}%
and six particles \cite{Leib05,Lu07} and to a more complex architecture \cite
{Hald07}.

The innovative character of the present work is enlightened by the
diagrams reported in Figure 1. While, according to the 1935 proposal the
nonlocal correlations were conceived to connect the dynamics of two
''microscopic'' objects, i.e. two spins within the well known EPR-Bohm
scheme here represented by diagram (a)\cite{Kwia95}, in the present work the
entanglement is established between a ''microscopic'' and a ''macroscopic'',
i.e. multi-particle quantum object, via cloning amplification: diagram (b).
The amplification is achieved by adopting a high-gain nonlinear (NL)
parametric amplifier acting on a single-photon input carrier of quantum
information, i.e., a qubit state: $\left| \phi \right\rangle $. This
process, referred to as ''quantum injected optical parametric
amplification'' (\textrm{QI-OPA})\ \cite{DeMa98,DeMa05b} turned out to be
particularly fruitful in the recent past to gain insight into several little
explored albeit fundamental, modern aspects of quantum information, as
quantum cloning machines \cite{DeMa05b,Pell03,DeMa05}, quantum U-NOT gate
\cite{DeMa02}, quantum no-signaling \cite{DeAn07}.
Here, by exploiting the amplification process, we convert a single photon
qubit into a Macro-qubit involving a large number of photons. Let us venture
in a more detailed account of our endeavor.

\begin{figure}[h]
\includegraphics[scale=.4]{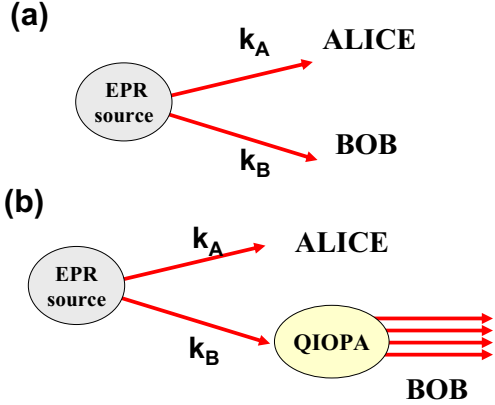}
\caption{(a) Generation of an entangled photon pair by Spontaneous
Parametric Down Conversion (SPDC)\ in a nonlinear (NL)\ crystal; (b)
Schematic diagram showing the single photon Quantum-Injected Optical
Parametric Amplification (QI-OPA).}
\end{figure}

\section{Test of Micro-Macro entanglement}

An entangled pair of two photons in the singlet state $\left| \Psi
^{-}\right\rangle _{A,B}$=$2^{-{\frac{1}{2}}}\left( \left| H\right\rangle
_{A}\left| V\right\rangle _{B}-\left| V\right\rangle _{A}\left|
H\right\rangle _{B}\right) \ $was produced through a Spontaneous Parametric
Down-Conversion (SPDC)\ by the NL\ crystal 1 (C1)\ pumped by a pulsed UV
pump beam: Fig.2. There $\left| H\right\rangle $ and $\left| V\right\rangle $
stands, respectively, for a single photon with horizontal and vertical
polarization while the labels $A,B$ refer to particles associated
respectively with the spatial modes $\mathbf{k}_{A}$and $\mathbf{k}_{B}$.
Precisely, $A,B$ represent the two space-like separated Hilbert spaces
coupled by the entanglement. The photon belonging to $\mathbf{k}_{B}$,
together with a strong ultra-violet (UV) pump laser beam, was fed into an
optical parametric amplifier consisting of a NL crystal 2 (C2)\ pumped by
the beam $\mathbf{k}_{P}^{\prime }$. More details on this setup and on its
properties are given in the Appendix and in \cite{Naga07}.
The crystal 2, cut for collinear operation, emitted over the two modes of
linear polarization, respectively horizontal and vertical associated with $%
\mathbf{k}_{B}$. The interaction Hamiltonian of the parametric amplification
$\widehat{H}=i\chi \hbar \widehat{a}_{H}^{\dagger }\widehat{a}_{V}^{\dagger
}+h.c.$ acts on the single spatial mode $\mathbf{k}_{B}$ where $\widehat{a}%
_{\pi }^{\dagger }$ is the one photon creation operator associated with the
polarization $\overrightarrow{\pi }$. The main feature of this Hamiltonian
is its property of ''phase-covariance'' for ''equatorial'' qubits $\left|
\phi \right\rangle $, i.e. representing equatorial states of polarization, $%
\overrightarrow{\pi }_{\phi }=2^{-1/2}\left( \overrightarrow{\pi }%
_{H}+e^{i\phi }\overrightarrow{\pi }_{V}\right) ,\overrightarrow{\pi }_{\phi
\perp }=\overrightarrow{\pi }_{\phi }^{\perp }$, in a Poincar\'{e} sphere
representation having $\overrightarrow{\pi }_{H}$ and $\overrightarrow{\pi }%
_{V}$ as the opposite ''poles'' \cite{Naga07}. The equatorial qubits are
expressed in terms of a single phase $\phi \in (0,2\pi )$ in the basis $%
\left\{ \left| H\right\rangle ,\left| V\right\rangle \right\} $. Owing to
the corresponding invariance under $U(1)$ transformations, we can then
re-write: $\widehat{H}$ =$\frac{1}{2}i\chi \hbar e^{-i\phi }\left( \widehat{a%
}_{\phi }^{\dagger 2}-e^{i2\phi }\widehat{a}_{\phi \perp }^{\dagger
2}\right) +h.c.$ where $\widehat{a}_{\phi }^{\dagger }=2^{-1/2}(\widehat{a}%
_{H}^{\dagger }+e^{i\phi }\widehat{a}_{V}^{\dagger })$ and $\widehat{a}%
_{\phi \perp }^{\dagger }=2^{-1/2}(-e^{-i\phi }\widehat{a}_{H}^{\dagger }+%
\widehat{a}_{V}^{\dagger })$. The generic polarization state of the injected
photon $\left| \psi \right\rangle _{B}$state on mode $\mathbf{k}_{B}$
evolves into the output state $\left| \Phi ^{\psi }\right\rangle _{B}=%
\widehat{U}\left| \psi \right\rangle _{B}$\ according to the OPA unitary
transformation $\widehat{U}$ \cite{Naga07}. The overall output state
amplified by the OPA apparatus is expressed, in any polarization equatorial
basis $\left\{ \overrightarrow{\pi }_{\phi },\overrightarrow{\pi }_{\phi
\perp }\right\} $, by the Micro-Macro entangled state commonly referred to
in the literature as the ''Schroedinger Cat State'' \cite{Schl01}

\begin{equation}
\left| \Sigma \right\rangle _{A,B}=2^{-1/2}\left( \left| \Phi ^{\phi
}\right\rangle _{B}\left| 1\phi ^{\perp }\right\rangle _{A}-\left| \Phi
^{\phi \perp }\right\rangle _{B}\left| 1\phi \right\rangle _{A}\right)
\label{outputstate}
\end{equation}
where the mutually orthogonal multi-particle ''Macro-states'' are: {\small
\begin{eqnarray}
\left| \Phi ^{\phi }\right\rangle _{B} &=&\sum\limits_{i,j=0}^{\infty
}\gamma _{ij}\frac{\sqrt{(1+2i)!(2j)!}}{i!j!}\left| (2i+1)\phi ;(2j)\phi
^{\perp }\right\rangle _{B} \\
\left| \Phi ^{\phi \perp }\right\rangle _{B} &=&\sum\limits_{i,j=0}^{\infty
}\gamma _{ij}\frac{\sqrt{(1+2i)!(2j)!}}{i!j!}\left| (2j)\phi ;(2i+1)\phi
^{\perp }\right\rangle _{B}
\end{eqnarray}
} with $\gamma _{ij}\equiv C^{-2}(-\frac{\Gamma }{2})^{i}\frac{\Gamma }{2}%
^{j}$, $C\equiv \cosh g$, $\Gamma \equiv \tanh g$, being $g$\ the NL\ gain%
\emph{\ } \cite{DeMa02}. There $\left| p\phi ;q\phi ^{\perp }\right\rangle
_{B}$ stands for a Fock state with $p$ photons with polarization $%
\overrightarrow{\pi }_{\phi }$ and $q$ photons with $\overrightarrow{\pi }%
_{\phi \perp }$ over the mode $\mathbf{k}_{B}$. Most important, any injected
single-particle qubit $(\alpha \left| \phi \right\rangle _{B}+\beta \left|
\phi ^{\bot }\right\rangle _{B})$ is transformed by the \textit{information
preserving} QI-OPA\ operation into a corresponding Macro-qubit $(\alpha
\left| \Phi ^{\phi }\right\rangle _{B}+\beta \left| \Phi ^{\phi \perp
}\right\rangle _{B})$, i.e. a ''Schr\oe dinger Cat'' like, macroscopic
quantum superposition \cite{DeMa98}. The quantum states of Eq.(2-3) deserve
some comments. The multi-particle states $\left| \Phi ^{\phi }\right\rangle
_{B}$, $\left| \Phi ^{\phi \perp }\right\rangle _{B}$ are orthonormal and
exhibit observables bearing macroscopically distinct average values.
Precisely, for the polarization mode $\overrightarrow{\pi }_{\phi }$ the
average number of photons is $\overline{m}=\sinh ^{2}g$ for $\left| \Phi
^{\phi \perp }\right\rangle_{B}$, and $(3\overline{m}+1)$ for $\left| \Phi
^{\phi }\right\rangle _{B}$. For the $\pi -$mode $\overrightarrow{\pi }%
_{\phi \perp }$ these values are interchanged among the two Macro-states. On
the other hand, as shown by \cite{DeMa98}, by changing the representation
basis from $\left\{ \overrightarrow{\pi }_{\phi },\overrightarrow{\pi }%
_{\phi \perp }\right\} \;$to $\left\{ \overrightarrow{\pi }_{H},%
\overrightarrow{\pi }_{V}\right\} $, the same Macro-states, $\left| \Phi
^{\phi }\right\rangle _{B}$ or $\left| \Phi ^{\phi \perp }\right\rangle _{B}$
are found to be quantum superpositions of two orthogonal states $\left| \Phi
^{H}\right\rangle _{B}$, $\left| \Phi ^{V}\right\rangle _{B}$ which differ
by a single quantum. This unexpected and quite peculiar combination, i.e. a
large difference of a measured observable when the states are expressed in
one basis and a small Hilbert-Schmidt distance of the same states when
expressed in another basis turned out to be a useful and lucky property
since it rendered the coherence patterns of our system very robust toward
coupling with environment, e.g. losses. This was verified experimentally.
The decoherence of our system was investigated experimentally and
theoretically in the laboratory: cfr: \cite{DeMa05,Cami06,Naga07}. The above
features are not present in atomic ensemble experiments, in which quantum
phenomena usually involve microscopic fluctuations of a macroscopic system
and the qubit states are encoded as collective spin excitations.

As shown in Figure 2, the single particle field on mode $\mathbf{k}_{A}$ was
analyzed in polarization through a Babinet-Soleil phase-shifter (PS), i.e. a
variable birefringent optical retarder, two waveplates $\left\{ \frac{%
\lambda }{4},\frac{\lambda }{2}\right\} $ and polarizing beam splitter (%
\text{PBS}). It was finally detected by two single-photon detectors $D_{A}$
and $D_{A\text{ }}^{\ast }$ (ALICE box). \ The multiphoton QI-OPA\ amplified
field associated with the mode $\mathbf{k}_{B}$ was sent, through a
single-mode optical fiber (SM),\ to a measurement apparatus consisting of a
set of \ waveplates $\left\{ \frac{\lambda }{4},\frac{\lambda }{2}\right\} $%
, a (PBS) and two photomultipliers (PM) $P_{B}$ and $P_{B}^{\ast }$ (BOB
box). The output signals of the PM's were analyzed by an ''orthogonality
filter'' (OF) that will be described shortly in this paper.\

\begin{figure*}[t]
\includegraphics[scale=.4]{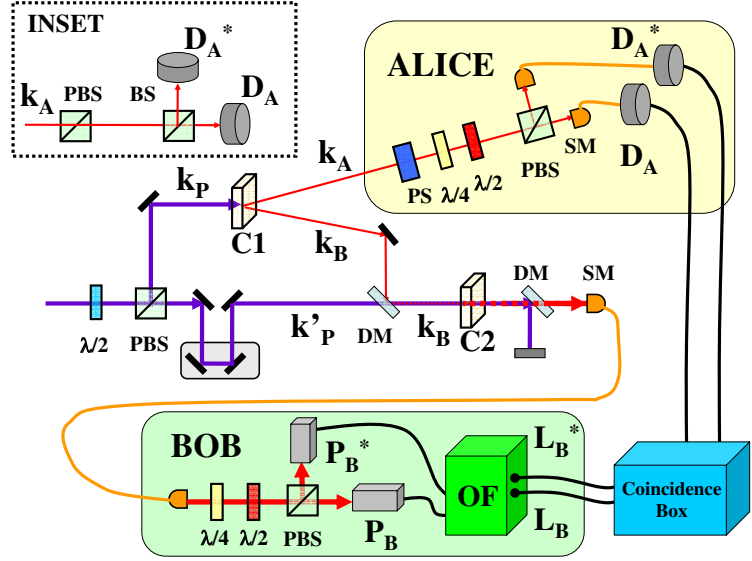}
\caption{Optical configuration of the QI-OPA apparatus. The SPDC quantum
injector (NL\ crystal 1: C1) is provided by a type II generator of
polarization-entangled photon couples. C1 generates an average photon number
per mode equal to about $0.35$, while the overall detection efficiency of
the trigger mode was estimated to be $\simeq 10\%.$ The NL crystal 2: C2,
realizing the optical parametric amplification (OPA), is cut for collinear
type II phase matching. Both crystals C1 and C2 are 1.5 mm thick. The fields
are coupled to single mode (SM) fibers. INSET: Two photon $\protect\pi -$%
states over the mode $\mathbf{k}_{A}$ are detected through a fiber coupled
beam-splitter $BS$ by a pair of single-photon detectors $D_{A}$ and $%
D_{A}^{\ast }$. A coincidence device, not shown in the Figure reveals a
coincidence between two simultaneous detection events by $D_{A}$ and $%
D_{A}^{\ast }$ with the emission of a transistor-transistor logic (TTL)
square signal $\overline{D_{A}}$ . \ For the amplification of the 2-photon
state, the intensity of the UV pump laser beam $\mathbf{k}_{P}$ was set at a
level apt to generate 3 pairs of photons in the NL crystal 1 with a low
probability: $p\sim 0.10$.}
\label{outputstate}
\end{figure*}

We now investigate the bipartite entanglement between the modes $\mathbf{k}%
_{A}$ and $\mathbf{k}_{B}$. We define the $\frac{1}{2}-$spin Pauli operators
$\left\{ \hat{\sigma}_{i}\right\} $ for a single photon polarization state,
where the label $i$ $=(1,2,3)$ refer to the polarization bases: $%
i=1\Longleftrightarrow \left\{ \overrightarrow{\pi }_{H},\overrightarrow{\pi
}_{V}\right\} $, $i=2\ \Longleftrightarrow \left\{ \overrightarrow{\pi }_{R},%
\overrightarrow{\pi }_{L}\right\} $, $i=3\Longleftrightarrow \left\{
\overrightarrow{\pi }_{+},\overrightarrow{\pi }_{-}\right\} $. Here $%
\overrightarrow{\pi }_{R}=2^{-1/2}(\overrightarrow{\pi }_{H}-i%
\overrightarrow{\pi }_{V}),\overrightarrow{\pi }_{L}=\overrightarrow{\pi }%
_{R}^{\perp }$ \ are the right and left handed circular polarizations and $%
\overrightarrow{\pi }_{\pm }=2^{-1/2}(\overrightarrow{\pi }_{H}\pm
\overrightarrow{\pi }_{V})$. It is found $\hat{\sigma}_{i}=\left| \psi
_{i}\right\rangle \left\langle \psi _{i}\right| -\left| \psi _{i}^{\perp
}\right\rangle \left\langle \psi _{i}^{\perp }\right| $ where $\left\{
\left| \psi _{i}\right\rangle ,\left| \psi _{i}^{\perp }\right\rangle
\right\} $ are the two orthogonal qubits corresponding to the $%
\overrightarrow{\pi }_{i}$ basis, e.g., $\left\{ \left| \psi
_{1}\right\rangle ,\left| \psi _{1}^{\perp }\right\rangle \right\} $ = $%
\left\{ \left| H\right\rangle ,\left| V\right\rangle \right\} $, etc. By the
QI-OPA unitary process the single-photon $\hat{\sigma}_{i}$ operators evolve
into the ''Macro-spin'' operators: $\hat{\Sigma}_{i}=\hat{U}\hat{\sigma}_{i}%
\hat{U}^{\dagger }=\left| \Phi ^{\psi i}\right\rangle \left\langle \Phi
^{\psi i}\right| -\left| \Phi ^{\psi i\perp }\right\rangle \left\langle \Phi
^{\psi i\perp }\right| .$ Since the operators $\left\{ \hat{\Sigma}%
_{i}\right\} $ are built from the unitary evolution of eigenstates of $\hat{%
\sigma}_{i}$ , they satisfy the same commutation rules of the single
particle $\frac{1}{2}-$spin: $\left[ \hat{\Sigma}_{i},\hat{\Sigma}_{j}\right]
=\varepsilon _{ijk}2i\hat{\Sigma}_{k}$ where $\varepsilon _{ijk}$ is the
Levi-Civita tensor density. The generic state $(\alpha \left| \Phi
^{H}\right\rangle _{B}+\beta \left| \Phi ^{V}\right\rangle _{B})$ is a
Macro-qubit in the Hilbert space $B$ spanned by $\left\{ \left| \Phi
^{H}\right\rangle _{B},\left| \Phi ^{V}\right\rangle _{B}\right\} $, as
said. To test whether the overall output state is entangled, one should
measure the correlation between the single photon spin operator $\hat{\sigma}%
_{i}^{A}$ on the mode $\mathbf{k}_{A}$ and the Macro-spin operator $\widehat{%
\Sigma }_{i}^{B}$ on the mode $\mathbf{k}_{B}$. We then adopt the criteria
for two qubit bipartite systems based on the spin-correlation. We define the
''visibility'' $V_{i}=\left| \left\langle \widehat{\Sigma }_{i}^{B}\otimes
\widehat{\sigma }_{i}^{A}\right\rangle \right| $ a parameter which
quantifies the correlation between the systems $A$ and $B.$ Precisely $%
V_{i}=\left| P(\psi _{i},\Phi ^{\psi i})+P(\psi _{i}^{\perp },\Phi ^{\psi
i\perp })-P(\psi _{i},\Phi ^{\psi i\perp })-P(\psi _{i}^{\perp },\Phi ^{\psi
i})\right| $ where $P(\psi _{i},\Phi ^{\psi i})$ is the probability to
detect the systems $A$ and $B$ in the states $\left| \psi _{i}\right\rangle
_{A}$ and $\left| \Phi ^{\psi i}\right\rangle _{B}$, respectively. The value
$V_{i}=1$ corresponds to perfect anti-correlation, while $V_{i}=0$ expresses
the absence of any correlation. The following upper bound criterion for a
separable state holds \cite{Eise04}:
\begin{equation}
S=(V_{1}+V_{2}+V_{3})\leq 1
\end{equation}
In order to measure the expectation value of $\widehat{\Sigma }_{i}^{B}$ a
discrimination among the pair of states $\left\{ \left| \Phi ^{\psi
i}\right\rangle _{B},\left| \Phi ^{\psi i\perp }\right\rangle _{B}\right\} $
for the three different polarization bases $1,2,3$ is required.\ Consider
the Macro-states $\left| \Phi ^{+}\right\rangle _{B}$, $\left| \Phi
^{-}\right\rangle _{B}$ expressed by Equations 2-3, for $\phi =0$\ and $\phi
=\pi $. In principle, a perfect discrimination can be achieved by
identifying whether the number of photons over the $\mathbf{k}_{B}$ mode
with polarization $\overrightarrow{\pi }_{+}$ is even or odd, i.e. by
measuring an appropriate ''parity operator''. This requires the detection of
the macroscopic field by a perfect \ \textit{photon-number resolving}
detectors operating with an overall quantum efficiency $\eta $ $\approx $ 1,
a device out of reach of the present technology.

\begin{figure}[h]
\includegraphics[scale=0.6]{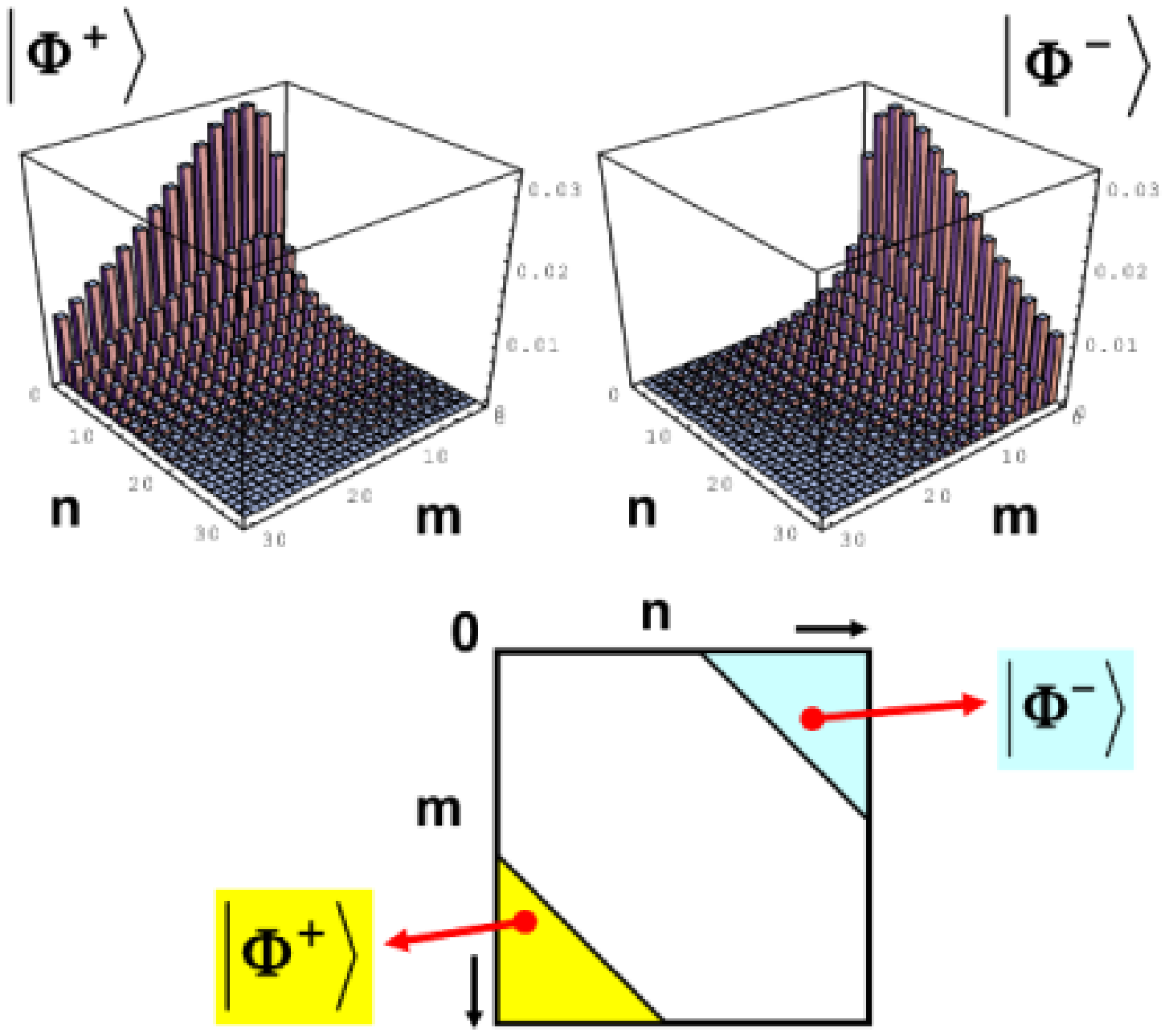}
\caption{Theoretical\ probability distributions $P^{\pm }(m,n)$ of the
number of photons associated with the Macro-states $\left| \Phi ^{\pm
}\right\rangle $ ($g=1.6$). Probabilistic identification of the
wavefunctions $\left| \Phi ^{\pm }\right\rangle $ by OF-filtering the $%
P^{\pm }(m,n)$ distributions over the photon number two-dimensional space $%
\left\{ m,n\right\} $. The white section in the cartesian plane $(m,n)$
corresponds to the ''inconclusive events'' of our POVM OF-filtering
technique.}
\end{figure}

It is nevertheless possible to exploit, by a somewhat sophisticated
electronic device dubbed ''Orthogonality Filter'' (OF), the macroscopic
difference existing between the functional characteristics of the
probability distributions of the photon numbers associated with the quantum
states $\left\{ \left| \Phi ^{\pm }\right\rangle _{B}\right\} $. The
measurement scheme works as follows: Figures 2 and 3. The multiphoton field
is detected by two PM's $(P_{B},P_{B}^{\ast })$ which provide the electronic
signals $(I_{+},I_{-})$ corresponding to the field intensity on the mode $%
\mathbf{k}_{B}$ associated with the $\pi -$components $(\overrightarrow{\pi }%
_{+},\overrightarrow{\pi }_{-})$, respectively. By (OF) the
difference signals $\pm (I_{+}-I_{-})$ are compared with a
threshold $\xi k>0$ . When the condition $(I_{+}-I_{-})>\xi k$
($(I_{-}-I_{+})>\xi k$) is satisfied, the detection of the state
$\left| \Phi ^{+}\right\rangle _{B}$ ($\left| \Phi
^{-}\right\rangle _{B}$) is inferred and a standard
transistor-transistor-logic (TTL) electronic square-pulse $L_{B}$
($L_{B}^{*}$) is realized at one of the two output ports of (OF).
The PM\ output signals are discarded for $-\xi k<(I_{+}-I_{-})<\xi
k$, i.e. in condition of low state discrimination. By increasing
the value of $\ $the\ threshold $k$ an increasingly better
discrimination is obtained together with a decrease of detection
efficiency. This ''local distillation'' procedure is conceptually
justified by the following theorem: since entanglement cannot be
created\ or enhanced by any ''local'' manipulation of the quantum
state, the non-separability condition demonstrated for a
''distilled'' quantum system, e.g., after application of the
OF-filtering procedure, fully applies to the same system in
absence of distillation \cite{Eise04}. This statement can be
applied to the measurement of $I_{\phi }$ and $I_{\phi \perp }$
for any pair of quantum states $\left\{ \left| \Phi ^{\phi
}\right\rangle _{B},\left| \Phi ^{\phi \perp }\right\rangle
_{B}\right\} $. This method is but an application of a Positive
Operator Value Measurement \ procedure (POVM)\ \cite{Pere95} by
which a large discrimination between the two states $\left\{
\left| \Phi ^{\pm }\right\rangle _{B}\right\} $ is attained at the
cost of a reduced probability of a successful detection.\ A
detailed description of\ the OF device and of its properties is
found in the Appendix and in Ref.\cite{Naga07}.

The present experiment was carried out with a gain value $g=4.4$ leading to
a number of output photons $N\approx 3\times 10^{4}$, after OF filtering. In
this case the probability of photon transmission through the OF\ filter was:
$p\approx 10^{-4}$. A NL gain $g=6$ was also achieved with no substantial
changes of the apparatus. Indeed, an unlimited number of photons could be
generated in principle by the QI-OPA technique, the only limitation being
due to the fracture of the NL crystal 2 in the focal region of the laser
pump. In order to verify the correlations existing between the single photon
generated by the NL\ crystal 1 and the corresponding amplified Macro-state,
we have recorded the coincidences between the single photon detector signal $%
D_{A}$ (or $D_{A}^{\ast }$) and the TTL signal $L_{B}$ (or $L_{B}^{\ast }$)
both detected in the same $\pi-$basis $\left\{ \overrightarrow{\pi }_{+},%
\overrightarrow{\pi }_{-}\right\} $:\ Figure 2. This measurement has\ been
repeated by adopting the common basis $\left\{ \overrightarrow{\pi }_{R},%
\overrightarrow{\pi _{L}}\right\} $.

Since the filtering technique can hardly be applied to the $\left\{
\overrightarrow{\pi }_{H},\overrightarrow{\pi }_{V}\right\} $ basis, because
of the lack of a broader$\ SU(2)$ covariance of the amplifier, the small
quantity $V_{1}>0$ could not be\ precisely measured. The phase$\ \phi $
between the $\pi -$components $\overrightarrow{\pi }_{H}$ and $%
\overrightarrow{\pi }_{V}$ on mode $\mathbf{k}_{A}$ was determined by the
Babinet-Soleil variable phase shifter ($PS$)$.$ Figure 4 shows the fringe
patterns obtained by recording the rate of coincidences of the signals
detected by the Alice's and Bob's measurement apparata,\ for different
values of $\phi $. These patterns were obtained by adopting the common
analysis basis $\left\{ \overrightarrow{\pi }_{R},\overrightarrow{\pi }%
_{L}\right\} \;$with a filtering probability $\simeq 10^{-4}$,
corresponding to a threshold $\xi k$ about $8$ times higher than
the average photomultiplier signals $I$. In this case the average
visibility has been found $V_{2}=(54.0\pm 0.7)\%$. A similar
oscillation pattern has been obtained in the basis $\left\{
\overrightarrow{\pi }_{+},\overrightarrow{\pi }_{-}\right\} $
leading to: $V_{3}=\left( 55\pm 1\right) \%$. Since always is
$V_{1}>0$, our experimental result $S\;$= $V_{2}+V_{3}\;$=
$(109.0\pm 1.2)\%$ implies the violation of the separability
criteria of Equation (4) and then demonstrates the
non-separability of our Micro-Macro system belonging to the
space-like separated Hilbert spaces $A$ and $B$. By evaluating\
the experimental value of the ''concurrence'' for our test,
connected with the ''entanglement of formation'', it is obtained
$C\geq 0.10\pm 0.02>0$ \cite{Benn96,Hill97}. This result again
confirms the non-separability of our bipartite system.\ Further
details on the measurement can be found in the Appendix. A method
similar to ours to test the non-separability of a 2-atom
bi-partite system was adopted recently by \cite{Moeh07}.

\begin{figure}[h]
\includegraphics[scale=.2]{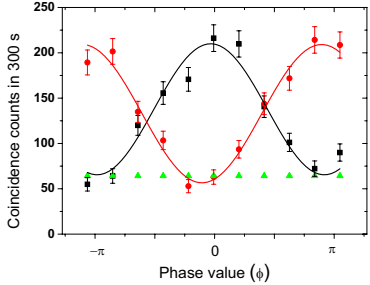}
\caption{Coincidence counts versus the phase $\protect\phi $ of the injected
qubit for a common basis $\left\{ \vec{\protect\pi }_{+},\vec{\protect\pi }%
_{-}\right\} $: square data $[L_{B},D_{A}]$, circle data $[L_{B}^{\ast
},D_{A}]$. The ''Visibility'' of the fringe pattern is: $V$ $\simeq 55\%$.
Triangle data: noise due to accidental coincidences.}
\end{figure}

\section{Violation of the Micro-Macro Bell's inequalities}

A further investigation on the persistence of quantum effects in large
multi-particle systems has been carried out by performing a nonlocality test
implying the violation of a Bell's inequality \cite{Reid02}. To carry out
such a test higher correlations among the fields realized on the $\mathbf{k}%
_{A}$ and $\mathbf{k}_{B}$ modes are required. Hence we adopted the QI-OPA\
amplification of \ a 4-photon entangled state. In this second experiment,
the NL crystal 1 generated by SPDC two simultaneous entangled photon
couples, i.e. a 4-photon entangled state . The 4 photons were emitted over
the two output modes $\mathbf{k}_{i}$ ($i=A,B$) in the \textit{singlet}
state of two spin-1 subsystems {\small
\begin{equation}
|\Psi _{2}^{-}\rangle_{AB}=\frac{1}{\sqrt{3}}(|2H\rangle _{A}|2V\rangle
_{B}-|H;V\rangle _{A}|H;V\rangle _{B}+|2V\rangle _{A}|2H\rangle _{B})
\end{equation}
} We note that the state $|\Psi _{2}^{-}\rangle_{AB}$ keeps the same
expression in any polarization basis owing to its rotational invariance. The
2 photons generated over the mode $\mathbf{k}_{B}$ were injected into the
collinear QI-OPA amplifier: Figure 2. After the amplification process the
overall output state can be expressed in any ''equatorial'' polarization
basis\ $\left\{ \overrightarrow{\pi }_{\phi },\overrightarrow{\pi }_{\phi
}^{\perp }\right\} \;$on \ the Poincar\'{e} sphere as {\small
\begin{equation}
\left| \Omega \right\rangle_{AB}=\frac{1}{\sqrt{3}}\left( |2\phi \rangle
_{A}\left| \Phi ^{2\phi ^{\perp }}\right\rangle _{B}-|\phi ;\phi ^{\perp
}\rangle _{A}\left| \Phi ^{\phi ,\phi ^{\perp }}\right\rangle _{B}+|2\phi
^{\perp }\rangle _{A}\left| \Phi ^{2\phi }\right\rangle _{B}\right)
\end{equation}
} where $\left| \Phi ^{\psi }\right\rangle _{B}$ stands for the amplified
field generated by injecting the 2-photon state $\left| \psi \right\rangle
_{B}$: $\left| \Phi ^{\psi }\right\rangle _{B}=\widehat{U}\left| \psi
\right\rangle _{B}$.

\begin{figure}[h]
\includegraphics[scale=0.6]{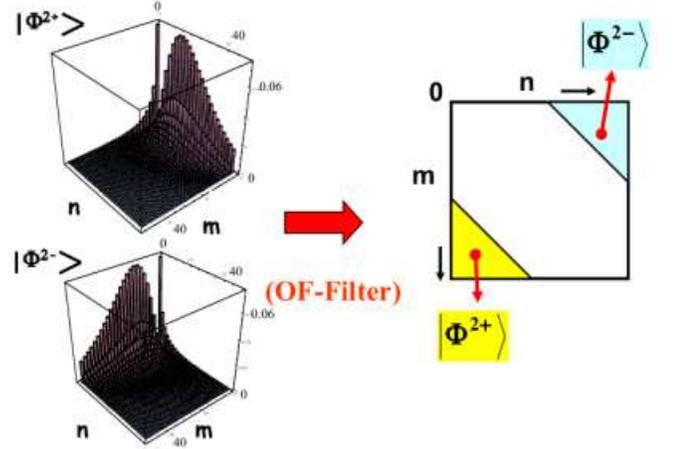}
\caption{Theoretical probability distribution $P^{\pm }(m,n)$ of the number
of photons associated with the Macro-states $\left| \Phi ^{2\pm
}\right\rangle $ ($g=1.6$). Probabilistic identification of wavefunctions $%
\left| \Phi ^{2\pm }\right\rangle $ by OF-filtering the $P^{\pm }(m,n)$
distributions over the photon number two-dimensional space $\left\{
m,n\right\} $.}
\end{figure}

Let us analyze the output field when the state $\left| 2\phi ^{\perp
}\right\rangle _{A}$ was detected over the mode $\mathbf{k}_{A}$. This was
done by the simultaneous measurement, by the single-photon detectors $D_{A}$
and $D_{A}^{\ast }\;$inserted\ on the two output modes of an ordinary
beam-splitter (BS),\ of two photons emerging simultaneously from the same
output port of a $\pi-$analyzer (\text{PBS),} inserted on the mode $\mathbf{k%
}_{A}\;$(see Figure 2-inset). Each event of simultaneous detection by $D_{A}$
and $D_{A}^{\ast }$was identified by the generation of a signal $\overline{%
D_{A}}$ \ at the output port of an additional electronic coincidence device,
not shown in Figure 2. In this condition the corresponding, correlated state
$\left| 2\phi \right\rangle _{B}$ was injected into the QI-OPA\ amplifier on
mode $\mathbf{k}_{B}$ leading to the generation of $\left| \Phi ^{2\phi
}\right\rangle _{B}$. When detected in the polarization basis $%
\overrightarrow{\pi }_{\pm }$ of the mode $\mathbf{k}_{B}$, the average
photon number $N_{\pm }$ was found to depend on the phase $\phi $ as
follows: $N_{\pm }(\phi )=\overline{m}+\left( 5\overline{m}+2\right) \cos
^{2}\left( \frac{\phi }{2}\right) $ with $\overline{m}=\sinh ^{2}g$. Hence
the output state analyzed in the polarization basis $\overrightarrow{\pi }%
_{\pm }$ exhibits a fringe pattern of the field intensity depending on $\phi
$ with a gain-dependent visibility $V_{th}$ $=(4\overline{m}+1)/(6\overline{m%
}+2)$.When $\phi =0$, the mode $\overrightarrow{\pi }_{+}$ is injected by a
two photon state, the output field is $\left| \Phi ^{2+}\right\rangle _{B}$
and $N_{+}(0)=5\overline{m}+2$. When $\phi =\pi ,$ the state $\left| \Phi
^{2-}\right\rangle _{B}$ is generated and a regime of spontaneous emission
is established: $N_{+}(\pi )=\overline{m}$.\ The two Macro-states $\left\{
\left| \Phi ^{2\pm }\right\rangle _{B}\right\} $ were singled out, in our
conditional experiment, by the simultaneous detection by the Alice's
Measurement apparatus of the correlated states $\left| 2\pm \right\rangle
_{A}$. As said, the realization of the state $\left| 2\phi \right\rangle _{A}
$ was identified by the emission of the signal $\overline{D_{A}}$ ,\ having
previously set the\ variable phase-shifter PS at the wanted phase value $%
\phi $. The Macro-states $|\Phi ^{2\pm }\rangle _{B}$ exhibited observables
macroscopically distinct, the difference being larger than the one observed
in the case of the QI-OPA\ amplification of a single photon state.
Accordingly,\ the $\left\{ \left| \Phi ^{2\pm }\right\rangle _{B}\right\} $
state distinguishability problem was found easier than in the single photon
case because of a lower mutual overlap of the 2-photon probability
distributions. The number of photons emitted by QI-OPA\ in this experiment,
after OF\ filtering, was $N\approx 3.5\times 10^{4}$, with a corresponding
transmission probability through the OF\ filter: $p\approx 10^{-3}$.
\begin{figure}[h]
\includegraphics[scale=.3]{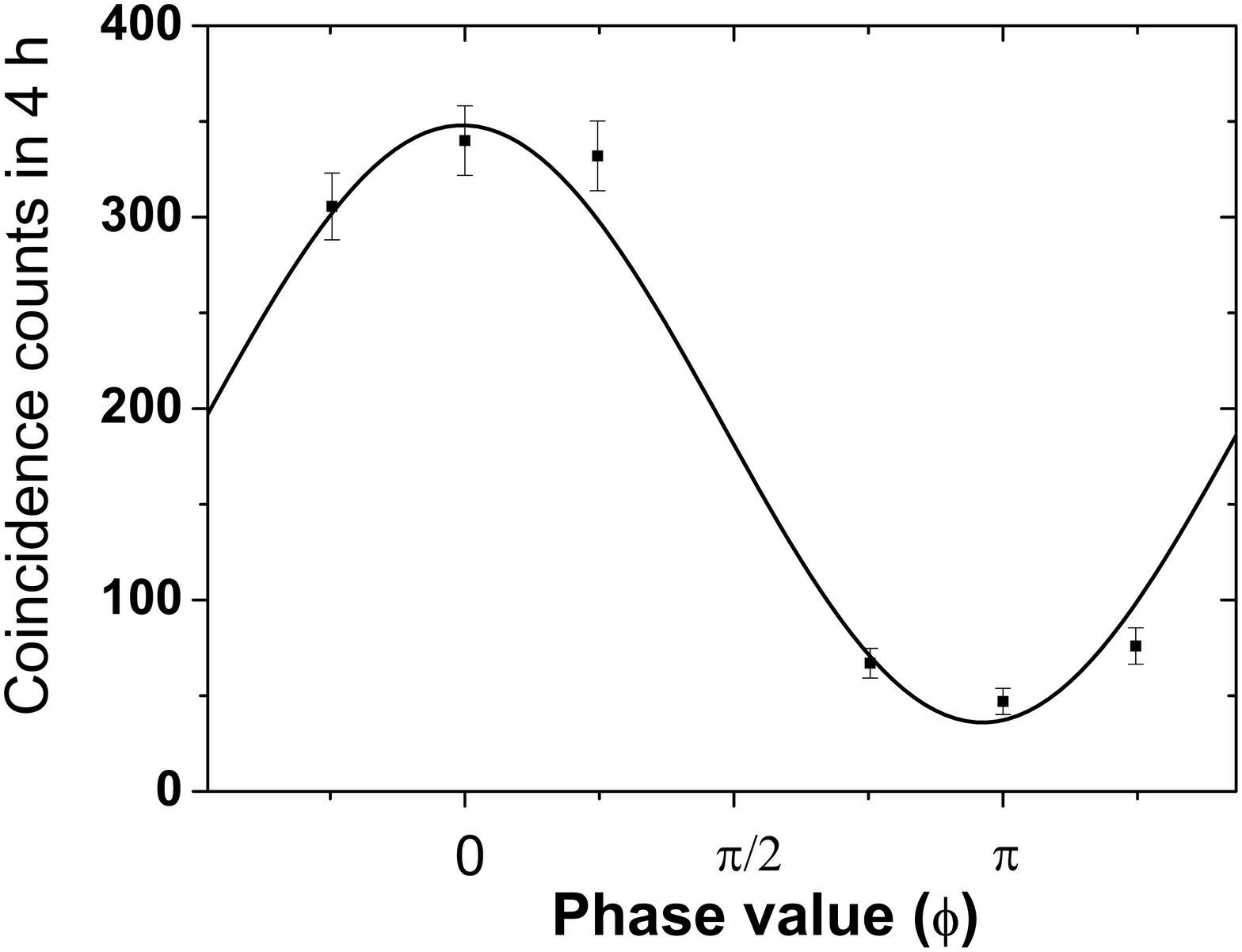}
\caption{Coincidence counts $[L_{B},\overline{D_{A}}]$ versus the phase $%
\protect\phi $ of the injected ''equatorial''qubit. ''Visibility'' of the
fringe pattern $V\simeq 81\%$.}
\end{figure}

We address now the problem of the distinguishability between
$\left| \Phi ^{2+}\right\rangle _{B}$ and $\left| \Phi
^{2-}\right\rangle _{B}$, which are mutually orthogonal as shown
by the quantum analysis reported in the Appendix. Figure 5 reports
the 3-dimensional representations of the two probability
distributions of the number of photons which are associated with
the two Macro-states $\left| \Phi ^{2\pm }\right\rangle _{B}.$
These distributions are drawn as functions of the variables $m$
and $n$ which are proportional to the size of the electronic
signals $I_{+}$ $=\xi m$ and $I_{-}=\xi n$ generated by the PM's $%
(P_{B},P_{B}^{\ast })$. In analogy with the entanglement experiment
accounted for in the previous section, a large discrimination between $%
\left\{ \left| \Phi ^{2\pm }\right\rangle _{B}\right\} $ could be obtained
by adopting the OF-filter. \ The output stations located in spaces $A$ and $%
B $ measured a dichotomic variable with eigenvalues $\pm 1$. The correlation
fringe patterns reported in Figure 6 were determined by simultaneous
detection on the mode $\mathbf{k}_{A}$ of the state $\left| 2\phi
\right\rangle _{A}$ for different values of the ''equatorial'' phase $\phi $
and on the mode $\mathbf{k}_{B}$ of the OF-filtered $\left\{ \left| \Phi
^{2\pm }\right\rangle _{B}\right\} $ . Precisely, this was done by recording
via the ''Coincidence Box'' the rate of coincidences between the signal $%
\overline{D_{A}}$ \ realized at Alice's site and the TTL signal realized at
one of the output ports of the OF-filter:\ Figure 2. The best-fit fringe
pattern reported in Figure 6 was obtained by recording, for different $\phi $
values the coincidence rate between $\overline{D_{A}}$ and the TTL signal
realized at the output port $L_{B}\;$of the OF-filter with a filtering
probability $p\approx $ $10^{-3}.$ The visibility of the fringe pattern was
found $V=(81\pm 2)\%$ , i.e. a large enough value that allows a Micro-Macro
non locality test for the entangled two spin-1 system. An identical but
complementary pattern, i.e. shifted by a phase $\phi =\pi ,$ was found by
collecting the coincidences between $\overline{D_{A}}$ and the TTL realized
at the port\ $L_{B}^{\ast }\;$of the OF-filter. This additional pattern is
not shown in Fig. 6.

Let us briefly outline the inequality introduced by Clauser, Horne, Shimony,
and Holt (CHSH) \cite{Bell65}. Each of two partners, $A$ (Alice) and $B\;$%
(Bob) measures a dichotomic observable among two possible ones, i.e. Alice
randomly measures either $\mathbf{a}$ or $\mathbf{a}^{\prime }$ while Bob
measures $\mathbf{b}$ or $\mathbf{b}^{\prime }$: Figure 7-\textbf{a.} For
any couple of measured observables $(A=\{\mathbf{a},\mathbf{a}^{\prime }\}$,
$B=\{\mathbf{b},\mathbf{b}^{\prime }\})$, we define the following
correlation function
\begin{equation}
E(A,B)=\frac{N(+,+)+N(-,-)-N(+,-)-N(-,+)}{N(+,+)+N(-,-)+N(+,-)+N(-,+)}
\end{equation}
where $N(\alpha ,\beta )$ stands for the number of events in which the
observables $A$ and $B$ have been found equal to the dichotomic outcomes $%
\alpha $ and $\beta $. Finally we define a parameter which takes into
account the correlations for the different observables.
\begin{equation}
S=E(\mathbf{a},\mathbf{b})+E(\mathbf{a}^{\prime },\mathbf{b})+E(\mathbf{a},%
\mathbf{b}^{\prime })-E(\mathbf{a}^{\prime },\mathbf{b}^{\prime })
\end{equation}
Assuming a local realistic theory CHSH\ found that the relation $\left|
S\right| \leq S_{CHSH}=2$ holds.

\begin{figure}[t]
\includegraphics[scale=.35] {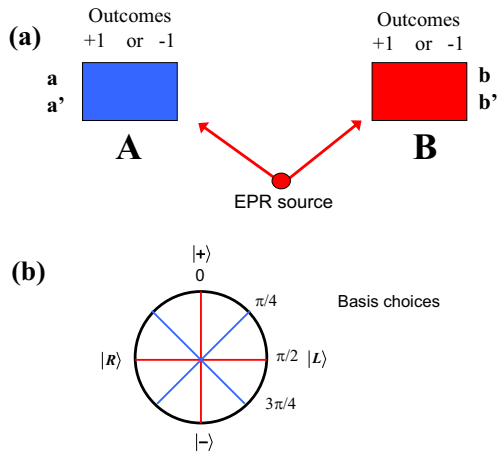}
\caption{(\textbf{a}) Conceptual scheme for testing the Bell-CHSH\
nonlocality Test. (\textbf{b}) Settings of the $\protect\phi $ phases by
which the Bell-CHSH test was carried out.}
\end{figure}

To carry out a non-locality test in the Micro-macro'' regime, we define the
two sets of dichotomic observables for A and B which can be measured over
different polarization basis sets. Both partners perform measurements in
equatorial basis $\left\{ \overrightarrow{\pi }_{\phi },\overrightarrow{\pi }%
_{\phi }^{\perp }\right\} $. A associates to the detection of the 2-photons
with polarization $\overrightarrow{\pi }_{\phi }$, i.e. of $\left| 2\phi
\right\rangle _{A}$, the observable $+1$ and to the detection of the
2-photons with polarization $\overrightarrow{\pi }_{\phi \perp }$, i.e. of $%
\left| 2\phi ^{\perp }\right\rangle _{A}$, the observable $-1.$ The two
possible output values $\{\mathbf{a},\mathbf{a}^{\prime }\}$ correspond to $%
\phi _{a}=\frac{\pi }{4}$ and $\phi _{a}^{\prime }=-\frac{\pi }{4}$: Figure
7-\textbf{b}$.$

B associates to the detection of the $\left| \Phi ^{2\phi }\right\rangle
_{B} $ the observable $+1$ and to the detection of the $\left| \Phi ^{2\phi
\perp }\right\rangle _{B}$ the observable $-1.$ The two possible values $\{%
\mathbf{b},\mathbf{b}^{\prime }\}$ correspond to $\phi _{b}=0$ and $\phi
_{b}^{\prime }=\frac{\pi }{2}.$ In order to achieve a high discrimination
among $\left\{ \left| \Phi ^{2\phi }\right\rangle _{B},\left| \Phi ^{2\phi
\perp }\right\rangle _{B}\right\} $ the OF-filter is adopted.\

Let us consider briefly the conceptual issues and the possible loopholes
raised by the present nonlocality test. As shown in Figure 7, our system\
fully reproduces the standard Bohm-Bell scheme for testing nonlocality for
an entangled pair of spin-1 particles \cite{Bouw02}. Two space-like
separated, uncorrelated ''black boxes'' $A$ and $B$ receive the entangled
photons from a common EPR source, i.e. involving the NL\ crystal 1 (C1).\
The box $A$ coincides with the\ ''ALICE Box'' in Figure 2 while the box $B$
contains all measurement devices appearing in the''BOB Box'' in Figure 2 and
the QI-OPA\ amplifier, i.e. the NL\ crystal 2 (C2). Assuming this
interpretation, the following two conditions characterize our present
nonlocality test: \ (1) No artificial selection, sampling or filtering
whatsoever is made on the photon couples emitted by the EPR source, before
the particles enter the boxes $A$ and $B$. (2)\ Any external operation
acting on the particles before measurement, e.g. amplification, loss due to
reduced quantum efficiencies, OF-filtering etc., is a ''local operation''
because it is exclusively\ attributable to the separated internal dynamics
of\ the devices contained in the boxes $A$ and $B$. On the basis of these
premises any sampling or filtering made on the\ particles by our system must
be defined a ''fair sampling'' operation \cite{Bell65}. There ''fairness''
is\ precisely implied by condition (1), i.e. stating that no externally
biased perturbation is allowed to act on the only carriers of nonlocality
connecting $A$ and $B$ : the EPR\ entangled particles. On these premises,
any hidden variable analysis of the overall process will possibly identify
new ''loopholes'' in our test. Certainly the simple ''detection loophole''
does not apply to our complex scheme \cite{Kwia94}. In this connection, we re%
$\min $d that several non-locality tests valid for post-selected events have
been conceived in the past. For instance, the ones adopting photonic\ GHZ
states \cite{Bouw99} or 4 photon cluster states \cite{Kies05}, in which the
tripartite entangled state is generated only when a detector fires on each
output mode. We believe that our test is similar to the last condition, the
only difference being due to the use made in either cases by the local
post-selection operation. While in the previous case it was instrumental to
generate a quantum state, in our case it is used to properly improve a local
measurement procedure, i.e. to make the quantum measurement sharper.\emph{\ }%
A procedure somewhat similar to ours was adopted by Babichev et al. to carry
out the non-locality proof of single photon dual-mode optical qubit \cite
{Babi04}. We believe that a careful quantum analysis will be required to
fully clarify the aspects of our POVM\ method in the context of any hidden
variable theory \cite{Pear70,Huel95}.

Experimentally we obtained the following values by carrying out a
measurement with a duration of $4$ hours and a statistics per setting equal
to about 500 events:
\begin{equation*}
\begin{tabular}{|c|c|c|c|}
\hline
$E(\phi _{a},\phi _{b})$ & $E(\phi _{a}^{\prime },\phi _{b})$ & $E(\phi
_{a},\phi _{b}^{\prime })$ & $E(\phi _{a}^{\prime },\phi _{b}^{\prime })$ \\
\hline
$0.643\pm 0.027$ & $0.551\pm 0.029$ & $0.608\pm 0.017$ & $-0.453\pm 0.023$
\\ \hline
\end{tabular}
\end{equation*}
which leads to
\begin{equation*}
S=2.256\pm 0.049
\end{equation*}
Hence a violation by more than $5$ standard deviation over the value $%
S_{CHSH}=2$ is obtained. This experimental value is in agreement with an
average experimental visibility of $V\sim 80\%$ which should lead to $%
S=2.26. $

\section{Applications to Quantum Information. Macro-Macro entanglement}

We have experimentally demonstrated the quantum non-separability of a
Micro-Macro-system. Furthermore, by increasing the size of the injected
''seed'' state, namely by adopting at the outset a 4-photon entangled state,
we have reported a violation of Bell inequalities. It is possible to
demonstrate that the methods adopted in present experiment can be
re-formulated in the context of the ''continuous variable'' approach (CV),
which is commonly used to analyze the quantum informational content of
''quadrature operators'' for multi-particle systems \cite{Brau05}. Indeed
the intersection of the present QI-OPA method and of the CV approach appears
to be an insightful and little explored field of research to which the
present work will contribute by eliciting an enlightening theoretical
endeavor.
However, in spite of these appealing
perspectives, we believe that the QI-OPA\ approach is more directly
applicable to the field of Quantum Information and Computation in virtue of
the intrinsic \textit{information-preserving} \ property of the QI-OPA\
dynamics. Indeed, this property implies the direct realization of the
quantum map $(\alpha \left| \phi \right\rangle +\beta \left| \phi ^{\bot
}\right\rangle )$ $\longrightarrow $ $(\alpha \left| \Phi ^{\phi
}\right\rangle +\beta \left| \Phi ^{\phi \perp }\right\rangle )$ connecting
any single-particle qubit to a corresponding Macro-qubit, by then allowing
the direct extension to the multi particle regime of most binary logic
methods and algorithms. A simple example may be offered by the
implementation of several universal 2-qubit logic gates such as the CNOT or
the phase-gate. Consider a 2-qubit phase gate in which the control-target
interaction is provided by a Kerr-type optical nonlinearity. It is well
known that the strength of this nonlinearity is far too small to provide a
sizable interaction between the''control'' and the ''target''
single-particle qubits, even in the special case of the atomic quasi
resonant electromagnetic induced transparency configuration (EIT) \cite
{Hau99,Flei05}. However, by replacing these qubits by the corresponding
Macro-qubits associated with $N$ photons, the NL interaction strength can be
enhanced by a large factor $\chi $ since the 3d-order NL\ polarization
scales as $N^{3/2}$. By assuming the values of the NL gain $g$ realized
experimentally in the present work, the value of this factor can be as large
as $\chi \approx 10^{7}$ for $g=4.4$\ and $\chi \approx 10^{11}$ for $g=6.0$%
. Such large enhancement of the NL interaction strength may
represent the key solution towards the technical implementation of
the most critical optical components of a quantum computer.\\
Of course, all these applications are made possible by an
important property of the QI-OPA scheme: there the multi-particle
entangled Macro-states are\ \textit{directly accessible}\ at the
output of the QI-OPA apparatus. In other words, in our systems the
many photons involved in the quantum superpositions are not sealed
or trapped in hardly accessible electromagnetic cavities, nor
suffer from decoherence processes due to handling, storing or
measurement procedures.
\begin{figure}[t]
\includegraphics[scale=.25]{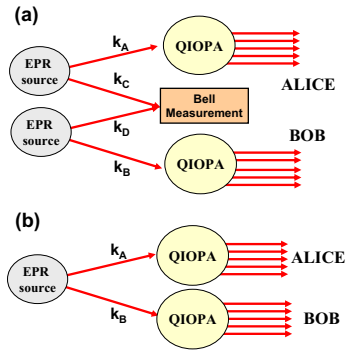}
\caption{Implementation of Macro-Macro entanglement (a)\ via two QI-OPA
entanglement swapping (b) through two optical parametric amplifiers.}
\end{figure}

At last, the Micro-Macro experimental method demonstrated in this work can
be upgraded in order to achieve an entangled Macro-Macro system showing
again marked nonlocality features. Such scheme could exploit an
''entanglement swapping'' protocol \cite{Pan98} as shown in Fig.8-(\textbf{a}%
). There the final entangled state is achieved through a standard
intermediate Bell measurement carried out on the Micro-states. A similar
process has been suggested in different contexts, for instance to entangle
micromechanical oscillators \cite{Pira06}. As an alternative approach, the
single photon states on mode $\mathbf{k}_{A}$ and $\mathbf{k}_{B}$ could be
amplified by two independent QI-OPA's :Fig.8-(\textbf{b}). Another appealing
perspective is the light-matter entanglement, consisting of the coupling,
either linear or nonlinear, of the multi-photon generated QI-OPA fields with
the mechanical motion of atomic systems. As an example, the coupling of a
Bose Einstein Condensed (BEC)\ assembly of \ Rb atoms with the entangled
multi-particle field generated by our apparatus is presently being
investigated in our Laboratories \cite{Cata07}. All this can open
interesting scenarios in modern science and technology. On a very
fundamental side, the observation of quantum phenomena with an increasing
number of particles can shed light on the elusive boundary between the
''classical'' and the ''quantum'' worlds, and provide new paths to
investigate the notion of quantum wavefunction\ and its intriguing
''collapse'' \cite{Guer07}.

\begin{acknowledgements}
We acknowledge technical support from Sandro Giacomini, and Giorgio Milani
and experimental collaboration with Eleonora Nagali,Tiziano De Angelis and
Nicolo' Spagnolo. This work was supported by the PRIN 2005 of MIUR and
project INNESCO 2006 of CNISM.
\end{acknowledgements}

\appendix
\section{Experimental setup}

We provide here more structural details on the apparatus shown in Figure 2
of the main text (MT) . The excitation source was a Ti:Sa Coherent MIRA
mode-locked laser amplified by a Ti:Sa regenerative REGA device operating
with pulse duration $180fs$, repetition rate $250kHz$ and average output
power: $1.6$ $W$. The pumping laser of the amplifier was a laser Coherent
Verdi operating at a continuous-wave $13.5W$ power level . The output beam,
frequency-doubled by second-harmonic generation, provided the OPA\
excitation field beam at the UV wavelength (wl) $\lambda _{P}=397.5nm$ with
power: $750\div 800mW$. The UV beam, splitted in two beams through a $%
\lambda /2$ waveplate (wp) and a polarizing beam splitter (PBS), excited two
BBO ($\beta $-barium borate) NL crystals cut for type II phase-matching.
Crystal 1, (C1) excited by the beam $\mathbf{k}_{P}$, was the SPDC source of
entangled photon couples with wl $\lambda =2\lambda _{P}$, emitted over the
two output modes $\mathbf{k}_{j}$ ($j=A,B$) in the \textit{singlet} state $%
\left| \Psi ^{-}\right\rangle _{A,B}$=$2^{-1/2}\left( \left| H\right\rangle
_{A}\left| V\right\rangle _{B}-\left| V\right\rangle _{A}\left|
H\right\rangle _{B}\right) \label{SPDCentangled}$. The power of beam $%
\mathbf{k}_{P}$ was set at a low enough level to generate with negligible
probability more than two simultaneous pairs of photons. The photon
associated with mode $\mathbf{k}_{A}$, hereafter referred to as \textit{%
trigger} mode, was coupled into a single mode fiber and excited through a
PBS\ one of the single photon counting modules (SPCM) $(D_{A},D_{A}^{\ast })$%
, while the single photon state generated over the mode $\mathbf{k}_{B}$ was
injected, together with a strong UV pump beam (mode $\mathbf{k}_{P}^{\prime }
$), into the NL\ crystal 2 (C2)\ and stimulated the simultaneous emission of
large number of photon pairs. The measurement was carried out on the mode $%
\mathbf{k}_{A}$ by adopting a set of two waveplates, $\lambda /2$ + $\lambda
/4$, a Soleil-Babinet variable phase-shifter ($PS$) and a polarizing
beam-splitter $PBS$. By a delay ($Z$) the time superposition in the OPA of
the excitation UV pulse and of the injection photon wavepacket was provided.
The injected single photon and the UV pump beam $\mathbf{k}_{P}^{\prime }$
were superimposed by means of a dichroic mirror ($DM$) with high
reflectivity at $\lambda $ and high transmittivity at $\lambda _{P}$. The
output state of the crystal $2$ with wl $\lambda $ was spatially separated
by the fundamental UV beam again by a dichroic mirror ($DM$)\ and spectrally
filtered by an interference filter ($IF$) with bandwidth $\Delta \lambda =$ $%
1.5nm$, transmittivity ($\approx 80\%$) and coupled to a single mode fiber ($%
SM$), polarization analyzed and then detected by 2 equal
photomultipliers (PM) $P_{B}$ and $P_{B}^{\ast }$ . The SPCM
detectors were single-photon modules Perkin Elmer type AQR14-FC.
The PM's were Burle A02 with a Ga-As photocathode having a
detector quantum efficiency $\eta _{QE}^{D}=13\%$.

In a first experiment with no quantum injection, we measured the gain value $%
g$ of the optical parametric process and the overall detection efficiency $%
\eta _{QE}$ of the detection apparatus. The average signal amplitude of $%
P_{B}$ was measured for different values of the UV power. The gain value $g$
of the process was obtained by fitting the experimental data \cite
{Eise04,Cami06} with an exponential function, leading to $g_{\max }=\left(
4.40\pm 0.02\right) $, corresponding to an overall mean photon number per
mode $\overline{m}=1470$. A gain $g=6.0$, $\overline{m}=4.0\times 10^{4}$
was also realized with no substantial changes in the apparatus. The
exponential growth demonstrates the multiple generation of photon pairs by a
self-stimulation process within the NL crystal. The overall efficiency on
the $\mathbf{k}_{B}$ mode, $\eta _{B}\simeq 3\%$ was determined by fiber
coupling ($\sim 50\%$), $IF$ transmittivity and detector $\eta _{QE}^{D}$.

\subsection{Orthogonality Filter}

\begin{figure}[b]
\vspace{1cm} \includegraphics[scale=.35]{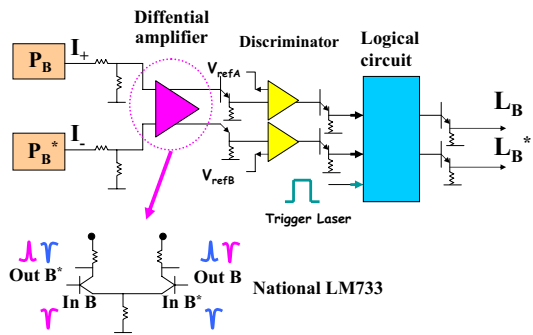} 
\caption{Electronic Orthogonality filter OF. The electronic signals $%
(I_{+},I_{-})$ emitted simultaneously by two photomultipliers $P_{B}$ and $%
P_{B}$ feed a linear difference amplifier (National LM733).\ Each
of the two output ports of the amplifier is connected to an
electronic discriminator set at a threshold level $\protect\xi
k>0$ equal for the two discriminators. Each discriminator emits a
TTL\ electronic square signal if the threshold is overcome by the
difference signals. Precisely, a TTL signal is realized at
the port $L_{B\;}$when: $(I_{+}-I_{-})>\protect\xi k$ \ or at the port $%
L_{B}^{\ast }$ when $(I_{-}-I_{+})>\protect\xi k$. The two
discriminators never fire simultaneously. The rejected events, for
$-\protect\xi k<(I_{+}-I_{-})<\protect\xi k$, correspond to the
''inconclusive outcomes'' of our generalized POVM measurement
technique.} \label{fig:counts}
\end{figure}

Here we give more details of the ''Orthogonality filter'' (OF), a device
adopted to discriminate among the quantum states $\left\{ \left| \Phi ^{\pm
}\right\rangle _{B}\right\} $. This operation is realized by exploiting the
different functional characteristics existing between the corresponding
photon number distribution patterns $P^{+}(m,n)$ and $P^{-}(m,n)$
corresponding to the Macro states expressed respectively by Equations(2-3)
of MT for $\phi =0$\ and $\phi =\pi $ :$\;\left| \Phi ^{+}\right\rangle _{B}$
and $\left| \Phi ^{-}\right\rangle _{B}$. Precisely, since the single-photon
resolving detection is beyond the reach of present technology and $\eta
_{QE}^{D}<1$, we may consider that the two Fock components of the Macro
states, $\left| (2i+1)+;(2j)-\right\rangle $ and $\left|
(2i)+;(2j+1)-\right\rangle $ belonging respectively to the expressions of $%
\left| \Phi ^{+}\right\rangle _{B}$ and $\left| \Phi ^{-}\right\rangle _{B}$%
, generate the same signals at the output of any couple of PM\ detectors: $%
(I_{+},I_{-})\propto (m\approx 2i,\;n\approx 2j)$. However a
discrimination may still be carried out efficiently between the
above orthogonal Macro states by exploiting the different
''shape'' of the probability distributions $P^{+}(m,n)$ and
$P^{-}(m,n)$ for large values of $\left| n-m\right| $. For this
purpose, we introduce an appropriate threshold $k>0$ for signal
discrimination.\ By analyzing the two probability distributions we
infer that when $(m-n)>k$\ the signals $(I_{+},I_{-})$\ can be
attributed to the state $\left| \Phi ^{+}\right\rangle _{B}$\ with
a fidelity that
increases with the value of $k$. In this case the measured eigenvalue of $%
\widehat{\Sigma }_{3}^{B}$ is found = +1.\ Likewise,\ when
$(n-m)>k$ the signals $(I_{+},I_{-})$\ is inferred to correspond
to the state $\left| \Phi ^{-}\right\rangle_{B} $ with
$\widehat{\Sigma }_{3}^{B}$ eigenvalue = -1. The events which
satisfy the inequality $\ -k<(m-n)<k$ \ are discarded since there\
the two distributions approximately overlap impairing a reliable
state discrimination. This technique can be adopted even with a
low value of detection efficiency ($\eta \simeq 10^{-2}$) since it
is found that the\ functional characteristics of the distributions
$P^{\pm }(m,n)$\ which are pertaining to the discrimination of the
states $\left| \Phi ^{\pm }\right\rangle _{B}$\ are preserved
under the signal propagation over a lossy channel. This important
point has been carefully investigated theoretically and
experimentally in our laboratory \cite{Naga07}.

The measurement scheme just described has been physically implemented by the
OF shown in Figure 9, an electronic device by which the pulse heights of
the couple of input signals $(I_{+},I_{-})$ provided by two PM's $%
(P_{B},P_{B}^{\ast })$ are summed with opposite signs by a balanced linear
amplifier $($\emph{LA) }with ''gain'' $G$ (chip National LM733). Each of the
two signals $\pm \lbrack G(I_{+}-I_{-})]\equiv \pm \lbrack G\xi (m-n)]$
realized at the two symmetric outputs of \emph{\ (LA)} feeds an independent
electronic discriminator (AD9696) set at a common threshold level $G\xi k$ .
Owing to previous considerations the two discri$\min $ators never fire
simultaneously and each of them provides, when activated,\ a standard
transistor-transistor-logic (TTL) square signal at its output port. As said,
when the condition $(I_{+}-I_{-})>\xi k$ , i.e. $(m-n)>k$ \ is satisfied, a
TTL signal $L_{B}$ is generated and then the realization of the state $%
\left| \Phi ^{+}\right\rangle _{B}$ is inferred. Likewise, when $%
(I_{-}-I_{+})>\xi k$ a TTL signal $L_{B}^{\ast }$ is generated\
and the realization of the state $\left| \Phi ^{-}\right\rangle
_{B}$ is inferred. The events that are discarded for: $-\xi
k<(I_{+}-I_{-})<\xi k$ correspond to the ''inconclusive'' outcomes
of any POVM \cite{Pere95}.

The OF device has been tested and characterized in condition of spontaneous
emission, i.e., in absence of any quantum injection into C2. In this
condition the output TTL signals $L_{B}$($L_{B}^{\ast }$) were measured by
sending only the signal $I_{+}$($I_{-}$) as input and by varying the
threshold $k$. In this regime the number of photons generated per mode
should exhibit a thermal probability distribution: $P(n)=\frac{\langle
n\rangle ^{n}}{(1+\langle n\rangle )^{n+1}}$ with $\langle n\rangle $\
average photon number per mode. Hence the probability to detect a signal
above the threshold $k$ is:$\;\Pi (k)=\sum_{n=k}^{\infty }P(n)=\left( \frac{%
\langle n\rangle }{(1+\langle n\rangle )}\right) ^{k}.$ We have
experimentally checked the dependence on the threshold $k$ of the number of
counts, which is expressed by $R\times \Pi (k)$,\ being $R$ the repetition
rate of the source. The experimental data shown by Figure 10 represent a
fair support of the expected exponential behavior.

\begin{figure}[h]
\vspace{1cm} \includegraphics[scale=.5]{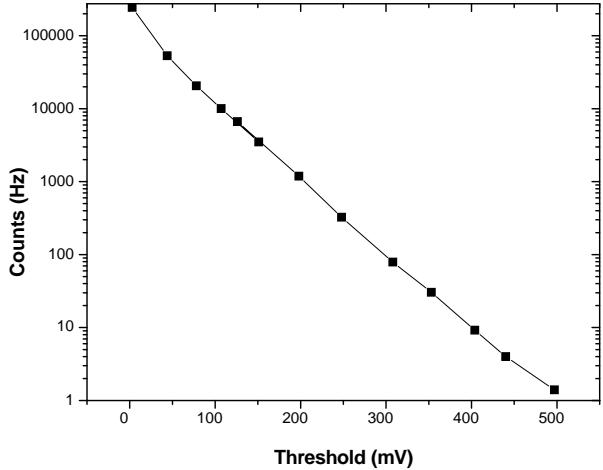} 
\caption{Count rates versus the threshold \emph{k}. The measured average
electronic signal was $\simeq 40mV$, while the repetition rate of the main
laser source was 250 kHz.}
\label{fig:counts}
\end{figure}

\subsection{Entanglement Test}

To carry out the measurement of the degree of the entanglement ALICE and BOB
adopt a common polarization basis $i$. In the basis $\left\{ \overrightarrow{%
\pi }_{+},\overrightarrow{\pi }_{-}\right\} $ the measurements
were carried out by setting the $\frac{\lambda }{4}$ and
$\frac{\lambda }{2}$ waveplates, shown in the ALICE Box in Figure
2, with the optical axes making angles
respect to the vertical direction equal to $45^{\circ }$ and $22.5^{\circ }$%
, respectively. An identical setting was adopted for the $\frac{\lambda }{4}$
and $\frac{\lambda }{2}$ wp's in the BOB Box in Figure 2. Likewise, in the
basis $\left\{ \overrightarrow{\pi }_{R},\overrightarrow{\pi }_{L}\right\} $%
\ the measurements were carried out by setting the two sets of $\frac{%
\lambda }{4},\frac{\lambda }{2}\;$wp's, shown in the ALICE and BOB Boxes,
with the optical axes making identical angles respect to the vertical $%
0^{\circ }$ and $22.5^{\circ }$, respectively. The phase shifter (\textrm{PS}%
), placed at the ALICE's site, consisted of \ a Soleil-Babinet compensator,
i.e. a variable birefringent retarder.

The experimental ''Visibility'' $V_{i}$ were measured by recording the
coincidence patterns $\{[L_{B},D_{A}],[L_{B}^{\ast
},D_{A}],[L_{B},D_{A}^{\ast }],[L_{B}^{\ast },D_{A}^{\ast }]\}$ recorded for
both spaces $A$ and $B$ in the basis $i$ \cite{Eise04}. According to the
standard definition: $V_{i}=(I_{\max }-I_{\min })/(I_{\max }+I_{\min })$
where $I_{\max },I_{\min }$ are respectively the maximum and the minimum
values of the detected signals corresponding to the value of phase $\phi =0$%
. The visibility values reported in the Main Text were recorded by averaging
over a time $\simeq 5$ hours the results of $4000$ events for each
experimental datum shown in Figure 4, for each basis.

It$^{\prime }$s worth noting that the above procedure is ''phase
covariant'', i.e. the overall quantum efficiency of the electronic filter OF
is independent of the phase $\phi $. In other words,$\ $and most impor$\tan $%
t in the present context, every state produced by QIOPA, $\left| \Phi ^{\phi
}\right\rangle _{B}=\widehat{U}\left| \phi \right\rangle _{B},$ has the same
overall probability to be filtered by the electronic filter and to produce a
TTL signal at one output port, either $L_{B}$ or $L_{B}^{\ast }$. This
feature was experimentally verified by measuring the OF-filtering
probability $P_{fil}$ for different phase values of the injected qubit. In
particular, $P_{fil}$ assumes the same value when the measurement is applied
to the state $\left\{ \left| \Phi ^{+}\right\rangle _{B},\left| \Phi
^{-}\right\rangle _{B}\right\} $ and $\left\{ \left| \Phi ^{R}\right\rangle
_{B},\left| \Phi ^{L}\right\rangle _{B}\right\} $ states.

\begin{figure}[t]
\includegraphics[scale=.8]{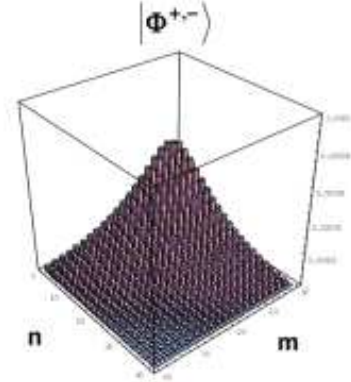}
\caption{Theoretical photon number probability distribution $P^{\pm }(m,n)$
for the Macro-state: $|\Phi ^{+,-}\rangle $ ($g=1.5$).}
\end{figure}

\subsection{CHSH tests}

Let us now account for the non-locality test by considering the measurement
carried out within the two Boxes A and B shown in Figure 7. (A) \textit{%
Measurement at the A Box}. The\ angles respect to the vertical made by the
optical axes of the waveplates $\frac{\lambda }{4},\frac{\lambda }{2}$ shown
in the ALICE Box in Figure 2 were set at $0^{\circ }$ and $22.5^{\circ }$,
respectively. The phase shifter (\textrm{PS}) was set in two different
positions to obtain $\phi _{a}=\frac{\pi }{4}$ and $\phi _{a}^{\prime }=-%
\frac{\pi }{4}$. (B) \textit{Measurement at the B Box}. The basis
corresponding to $\phi _{b}=0$ was obtained by setting the angles respect to
the vertical made by the optical axes of the wp's $\frac{\lambda }{4},\frac{%
\lambda }{2}$ shown in the BOB\ Box in Figure 2, at the values $45^{\circ }$
and $22.5^{\circ }$, respectively. The basis corresponding to $\phi
_{b}^{\prime }=\frac{\pi }{2}$ was obtained by setting the angles respect to
the vertical made by the optical axes of the same wp's $\frac{\lambda }{4},%
\frac{\lambda }{2}$ at the values $0^{\circ }$ and $22.5^{\circ
}$, respectively. As shown in Figure 2, in the B Box the above
phase changes were made by acting on the Macro-states, i.e., on
the output multi-particle beam emerging from the QI-OPA.

\begin{widetext}

\section{2-photon wavefunction}

The Macro-state generated by QI-OPA on mode $k_{B}$  when a 2-photon
state  $\left| (2)\pm \right\rangle _{B}$ is injected, is  expressed by

\begin{equation}
\left| \Phi ^{2\pm }\right\rangle _{B}=-\frac{\Gamma }{C\sqrt{2}}%
\sum\limits_{i,j=0}^{\infty }\gamma _{ij}\frac{\sqrt{(2i)!(2j)!}}{i!j!}%
\left| (2i)\pm ;(2j)\mp \right\rangle_{B} +\frac{1}{C^{3}}\sum\limits_{i,j=0}^{%
\infty }\gamma _{ij}\frac{\sqrt{(2i+2)!(2j)!}}{i!j!}\left|
(2i+2)\pm ;(2j)\mp \right\rangle_{B}
\end{equation}
While, the following Macro-state is generated if a state  $\left|
1+;1-\right\rangle$ is injected:

\begin{equation}
\left| \Phi ^{+,-}\right\rangle
_{B}=\frac{1}{C^{3}}\sum_{i,j}\left( -1\right) ^{j}\left(
\frac{\Gamma }{2}\right) ^{i+j}\frac{\sqrt{\left( 2i+1\right)
!\left( 2j+1\right) !}}{i!j!}\left|
(2i+1)+;(2j+1)-\right\rangle_{B}
\end{equation}

The photon number probability distribution for the Macro-state
$|\Phi ^{+,-}\rangle _{B} $   is reported in Figure 11.

\end{widetext}

\end{document}